\documentstyle[pre,aps,preprint,psfig]{revtex}     % journal page size

\newcommand{\figutwo}[4]{
                         \nobreak \hbox to \textwidth { \leavevmode 
                         \epsfxsize=#1 \epsffile{#2}
                         \hss \leavevmode 
                         \epsfxsize=#3 \epsffile{#4} }
                         \nobreak
                        }
\newcommand{\figu}[2]{\nobreak \hbox to \textwidth {
                      \centerline{\leavevmode
                      \epsfxsize=#1 \epsffile{#2}
                      }} \nobreak 
                     }

\newcommand{\bge}{\begin{equation}}
\newcommand{\bea}{\begin{eqnarray}}
\newcommand{\eea}{\end{eqnarray}}
\newcommand{\pde}{\partial}
\newcommand{\tr}{\mbox{Tr}}
\newcommand{\ede}{\end{equation}}
\newcommand{\ba}{\begin{array}}
\newcommand{\ea}{\end{array}}

\newcommand{\Lop}{{\cal L}}

\newcommand{\h}{\hbar}

\newcommand{\f}{\frac}

\newcommand{\p}{\partial}
\newcommand{\no}{\nonumber}
\newcommand{\field}{\phi}
\newcommand{\Lmat}{\mbox{\bf L}}
\newcommand{\Bmat}{\mbox{\bf B}}
\begin{document}
%\draft{
\title{Noise corrections to stochastic trace formulas}

\author{Gergely Palla and G\'abor Vattay}
\address{Department of Physics of Complex Systems, E{\"o}tv{\"o}s University\\
P\'azm\'any P{\'e}ter s{\'e}tany 1/A,
H-1117 Budapest, Hungary} 

\author{ Andr{\'e} Voros} 
\address{
CEA, Service de Physique Th{\'e}orique de Saclay\\
F-91191 Gif-sur-Yvette CEDEX, France\\
}
\author{Niels S\o ndergaard}
\address{
Department of Physics \&\ Astronomy, Northwestern University \\
2145 Sheridan Road, Evanston, IL 60208, USA}

\author{Carl Philip Dettmann}
\address{Department of Mathematics, University of Bristol\\
University Walk, Bristol BS8 1TW, United Kingdom}

\date{\today}

\maketitle

\begin{abstract}
We review studies of an evolution operator ${\cal L}$ for a discrete Langevin 
equation with a strongly hyperbolic 
classical dynamics and a Gaussian noise. The leading eigenvalue
of ${\cal L}$ yields a physically measurable property of the dynamical system,
the escape rate from the repeller.  The spectrum of the evolution 
operator ${\cal L}$ in the weak noise limit can be computed in several ways.
A method using a local matrix representation of the operator allows
to push the corrections to the escape rate up to order eight in the
 noise expansion parameter. These corrections then appear 
to form a divergent series. Actually, via a cumulant expansion,
they relate to analogous divergent series for other quantities,
 the traces of the evolution operators ${\cal L}^n$.
Using an integral
 representation of the evolution operator ${\cal L}$, we then investigate the
 high order corrections to the latter traces. Their asymptotic 
behavior is found to be controlled by sub-dominant saddle points 
previously neglected in the perturbative expansion, and to be ultimately
described by a kind of trace formula.

\end{abstract}

\section{Introduction}

In the statistical theory of dynamical systems the development of the
densities of particles is governed by a corresponding evolution
 operator.
For a repeller, the leading eigenvalue of this operator ${\cal L}$ 
yields a physically measurable property of the dynamical system,
the escape rate from the repeller.
In the case of deterministic flows, the periodic orbit theory \cite{A1,A2}
yields explicit and numerically efficient formulas for the spectrum of 
${\cal L}$ as zeros
of its spectral determinant \cite{QCcourse}.

Upon all dynamical evolutions in nature, stochastic 
processes of various strength have an influence.
In a series of papers \cite{noisy_Fred,conjug_Fred,diag_Fred,asymp}
 the effects of noise on measurable
properties such as dynamical averages in classical
 chaotic dynamical
systems were systematically accounted.
The theory developed is  closely related to the semi-classical $\h$ expansions
\cite{Gaspard,Vattay1,Vattay2}
 based
on Gutzwiller's formula for the trace in terms of classical periodic
orbits \cite{gutbook}, in that both are perturbative theories in the noise
strength or $\hbar$, derived from saddle-point expansions of a path
integral containing a dense set of unstable stationary points.
  The analogy with quantum mechanics and
field theory was made explicit in \cite{noisy_Fred} where
Feynman diagrams were used to find the lowest nontrivial
 noise corrections
 to the escape rate.

An elegant method, inspired by the classical perturbation theory of
celestial mechanics, was that of smooth conjugations \cite{conjug_Fred}.
In this approach
the neighborhood of each saddle point was flattened by an appropriate
coordinate transformation, so the focus shifted from the
 original dynamics
to the properties of the transformations involved. 
The expressions obtained for perturbative corrections in this approach
were much simpler than those found
from the equivalent Feynman diagrams.  Using these techniques, we were
able to extend the stochastic perturbation theory to the fourth order
 in the
noise strength.

In \cite{diag_Fred}  we developed
a third approach, based on construction of an explicit matrix
representation of the stochastic evolution operator. 
The numerical implementation required a
truncation to finite dimensional matrices,
and was less elegant than the smooth conjugation method,
but made it possible to reach  up to order eight in expansion orders.
As with the previous formulations, it retained the periodic orbit structure,
thus inheriting valuable information about the dynamics.

The corrections to the escape rate appeared to be a divergent series in the 
noise expansion parameter. They actually followed, via the
cumulant expansion, from the traces of 
the evolution operator ${\cal L}$ \cite{diag_Fred},
for which the noise correction series looked similarly divergent.
In \cite{asymp} the  high order corrections were then worked out analytically
for the special case of the first trace,  Tr$({\cal L})$,
as contour integrals asymptotically evaluated by the method of
 steepest descent.
Here we further confirm the divergent nature of the high order noise corrections
for Tr${\cal L}^n$ ; their high order terms are also controlled by sub-dominant
 saddles, which can be interpreted as generalized periodic orbits of some
 associated discrete Newtonian equations of motion. 

In the following sections, first we define
 the stochastic dynamics, the evolution operator and its spectrum.
 Since this work was inspired by the results obtained for the
eigenvalue corrections via the matrix representation method
 outlined in \cite{diag_Fred},
for the sake of completeness we next show how to obtain 
 a matrix representation of the evolution operator
 as an expansion in terms of the noise strength $\sigma$, and how to
 calculate the spectrum of the operator. Then we turn towards investigating
 the behavior of the high order noise corrections to the traces of
 $\Lop^n$, which are responsible for the divergent behavior seen
 at late terms of the eigenvalue corrections.
 Our key result is ($\ref{result2}$)
 where 
the high order noise corrections are converted into a trace formula. 
We give as a numerical example the quartic map considered
 in \cite{noisy_Fred,conjug_Fred,diag_Fred,asymp}.

\section{The stochastic evolution operator}
In this section we introduce the noisy repeller and its evolution operator. 
An individual trajectory in presence of additive noise is generated
by iterating 
\bea
x_{n+1}=f(x_{n})+\sigma\xi_{n} 
\eea
%\,,
%\ee{mapf(x)-Diag}
where $f(x)$ is a map, 
$\xi_n$ a random variable with the normalized
distribution $P(\xi)$, 
and $\sigma$ parameterizes the noise strength.
In what follows we shall assume that the mapping $f(x)$ is
one-dimensional and expanding, and that the $\xi_n$ are uncorrelated.
A density of trajectories $\phi(x)$ evolves with time on the average as
\bea
\phi_{n+1}(y) =
\left(
{\cal L}
\circ
\phi_{n}\right)(y)
= \int dx \, {\cal L}(y,x) \phi_{n}(x)
\label{DensEvol}
\eea
where the evolution operator ${\cal L}$ has the general form
\bea
{\cal L}(x,y) &=& \delta_\sigma(y-f(x)) ,\label{OpOverNoise} \\
  \delta_\sigma(x)
  &=& \int \delta(x-\sigma \xi) P(\xi) d\xi 
  \,=\, \frac{1}{\sigma} P\left( \frac{x}{\sigma} \right)
\,.
\label{oper-Diag}
\eea
 For the calculations
in this paper, Gaussian weak noise is assumed. The evolution operator
\bea
{\cal L}(x,y)=\f{1}{\sqrt{2\pi}\sigma}e^{-\f{(y-f(x))^2}{2\sigma^2}} 
\eea
can be expanded, in the perturbative limit $\sigma \rightarrow 0$, as
\bea
{\cal L}(x,y)\sim \sum_{N=0}^{\infty}a_{2N}\sigma^{2N}\delta^{(2N)}(y-f(x)),
\eea
where $\delta^{(2N)}$ denotes the $2N$-th derivative of the delta distribution
 and
$a_{2N}=(2^NN!)^{-1}$ (or $a_{2N}= \f{m_{2 N}}{(2 N)!}$ for general noise with 
finite $n$-th moment $m_n$).
The map used for concrete calculations is the same as in our previous papers, 
a quartic map on the $(0,1)$ interval given by
\bea
\label{map}
f(x)=20\left[\f{1}{16}-\left(\f{1}{2}-x\right)^4\right],
\eea
which is also shown on Fig. ($\ref{figmap}$).

Throughout the theory developed in previous 
works \cite{noisy_Fred,conjug_Fred,diag_Fred,asymp}, 
the periodic orbits of the system played a major role. A 
periodic orbit of length $n$ was defined simply by
\bea
x_{j+1}&=&f(x_j),\mbox{\hspace{1cm}}j=1,...,n\label{peri1}\\
x_{n+1}&\equiv&x_{1}. \label{peri2}
\eea
(This subscript $j$ will quite generally be defined mod $n$.)

For a repeller the leading eigenvalue of the evolution operator
yields a physically measurable property of the dynamical system,
the escape rate from the repeller.
In the case of deterministic flows, the periodic orbit theory
yields explicit formulas for the spectrum of ${\cal L}$ as zeros
of its spectral determinant \cite{QCcourse}. Our goal here
is to explore the dependence of the eigenvalues 
$\nu$ of ${\cal L}$ 
on the noise strength parameter $\sigma$.

The eigenvalues are determined by the eigenvalue condition
\bea
F(\sigma,\nu(\sigma))=\det(1-{\cal L}/\nu(\sigma)) =0
\label{eigCond}
\eea 
where
$
F(\sigma,1/z) = \det(1-z{\cal L})
%\label{SpecDet}
$
is the spectral determinant of  the evolution operator $\Lop$.
Computation of such determinants
starts with evaluation of the traces of powers of
the evolution operator
\bea
			         %   \left. 
\mbox{Tr} {z\Lop \over 1-z\Lop}        %\right|_\nCutoff
        &=&
         \sum_{n=1}^{\infty} C_{n} z^n
%         \sum_{n=1}^{\nCutoff} C_{n} z^n
        \,, \qquad C_{n} = \mbox{Tr} \Lop^n
\,,
\label{tr-L-ith-Diag}
\eea
which are then used to compute the cumulants
$Q_{n}=Q_{n}(\Lop)$ in  the cumulant expansion
\bea
			         %\left.
\det(1-z\Lop)			 %\right|_{\nCutoff} 
= 1- \sum_{n=1}^{\infty} Q_{n} z^n
%= 1- \sum_{n=1}^{\nCutoff} Q_{n} z^n
%\,,\qquad Q_{n}=Q_{n}(\Lop)= \mbox{$n$th cumulant}
\,,
\label{Fred-cyc-exp-Diag}
\eea
by means of the recursion formula
\bea
Q_n = {1 \over n}\left( C_n -C_{n-1} Q_1 - \cdots C_1 Q_{n-1}\right)
\label{Fd-cyc-exp-Diag}
\eea
which follows from the identity
%$\log\det=\tr\log$
\bea
%       F(z)
        \det(1-z\Lop)
        =
        \exp\left(-\sum_n^\infty {z^n \over n} \mbox{Tr} \Lop^n \right)
\,.
\label{det-tr-Diag}
\eea
In the next section we show how to compute the cumulants $Q_n$ using a 
local matrix representation of the evolution operator.

\section{The spectrum of the evolution operator}
\label{Matrix}

As the mapping $f(x)$ is expanding by assumption,  the evolution operator 
($\ref{DensEvol}$)  smoothes the initial
distribution $\phi(x)$. Hence it is natural to assume that
the distribution $\phi_{n}(x)$ is analytic, and represent it as
a Taylor series, intuition being that the action of $\Lop$ will 
smooth out fine detail in initial distributions and the expansion
of $\phi_{n}(x)$ will be dominated by the leading terms in the series.

An analytic function $g(x)$ has a Taylor series expansion
\[
g(x) = \sum_{m=0}^\infty
  \frac{x^{m}}{m!} \left.  \frac{\pde^{m}}{\pde y^{m}}g(y) \right|_{y=0}
\,.  
\]
Following H.H.~Rugh \cite{Rugh92} we
now define the matrix ($m,m'= 0,1,2, ...$)
\bea
 \left(\Lmat{}\right)_{m'm} = 
 \left. \frac{\pde^{m'}}{\pde y^{m'}}
  \int dx \, \Lop(y,x)
\frac{x^{m}}{m!} \right|_{y=0} .  
\label{Lmat-Diag}
\eea
$ \Lmat{}$ is a matrix representation of $\Lop$ which 
maps the $x^m$ component of the density of trajectories $\phi_n(x)$ in
 ($\ref{DensEvol}$) to the  $y^{m'}$ component of the density $\phi_{n+1}(y)$, with $y=f(x)$.
The desired traces can now be evaluated as traces of the matrix 
representation $ \Lmat{}$,
$
\tr \Lop^n
	=
\tr \Lmat{}^n
\,.
$
As $\Lmat{}$ is infinite dimensional, in actual
computations we have to truncate it to a given finite order.
For expanding flows the structure of $ \Lmat{}$
is such that its finite truncations give very accurate spectra.

Our next task is to evaluate the matrix elements of $ \Lmat{}$.
Traces of powers of the evolution operator $\Lop^n$ are now also a power
series in $\sigma$, with contributions composed of 
$\delta^{(m)}(f(x_a)-x_{a+1})$ segments.
The contribution is non-vanishing only if the sequence
$x_1,x_2,...,x_n, x_{n+1} = x_{1}$ is a periodic orbit of
the deterministic map $f(x)$.
Thus  the series expansion of $\tr \Lop^n$ has support on
all periodic points $x_a = x_{a+n}$ of period $n$,
$f^n(x_a)=x_a$; the skeleton of periodic points of the deterministic
problem also serves to describe the weakly stochastic flows.
The contribution of the
$x_a$ neighborhood 
is best presented by introducing a coordinate system $\field_a$
centered on the  cycle points,
 and the operator (\ref{OpOverNoise})
centered on the
$a$-th cycle point
\bea
x_a &\to& x_a+\field_a \,, \qquad a=1,...,n_p\\	
f_a(\field) &=& f(x_a+\field) \\ 
{\Lop}_a(\field_{a+1},\field_a)
	    &=&
   \Lop(x_{a+1}+\field_{a+1},x_a+\field_a)
\,.
\eea
The weak noise expansion  for the $a$-th segment operator
is given by
\bea
{\Lop}_a(\field',\field)=\sum_{N=0}^{\infty}(- \sigma)^{2N}
a_{2N} \delta^{(2N)}(\field'+x_{a+1}- f_a(\field))
\,.
%\label{li}
\eea

Repeating the steps that led to ($\ref{Lmat-Diag}$)
we construct the local matrix representation of $\Lop_{a}$ centered on
the $x_a \to x_{a+1}$ segment of the deterministic trajectory
\bea
 \left(\Lmat{a}\right)_{m'm} 
	&=& 
 \left. \frac{\pde^{m'}}{\pde \field'^{m'}}
  \int d\field \,\Lop_{a}(\field',\field)
\frac{\field^{m}}{m!} \right|_{\field'=0} .  \\
	&=& 
%(\Lmat{a})_{ k' k} 
 \sum_{N=max(m-m',0)}^{\infty}(-\sigma)^{2N}a_{2N}(\Bmat{a})_{m'+N,m}
\,.
\label{BtoL}
\eea
Due to its simple dependence on the Dirac delta function,
$\Bmat{a}$ can expressed in terms of derivatives of the inverse
 of $f_a(\field)$:
\bea
(\Bmat{a})_{n m}
	&=& 
	\left. \frac{\pde^{n}}{\pde \field'^{n}}
    \int d\field \, 
     \delta(\field' + x_{a+1} -f_a(\field))
    \frac{\field^{m}}{m!}
    \right|_{\field'=0}\\
	&=& 
   \left. \frac{\pde^{n}}{\pde \field'^{n}} \frac{(f_a^{-1}(x_{a+1}+\field')-x_a)^{m}}{m!|f_a'(f_a^{-1}(x_{a+1}+\field'))|}\right|_{\field'=0}
        \\
        &=&
   \frac{\mbox{sign}(f_a')}{(m+1)!}\left.\frac{\pde^{n+1}(
{\cal F}_a(\field')^{m+1})}{\pde \field'^{n+1}}\right|_{\field'=0},  
\label{Bmatrix}
\eea
 where we introduced the shorthand notation ${\cal F}_a(\field')=
f_a^{-1}(x_{a+1}+\field')-x_a$.
The matrix elements can be easily worked out explicitly using 
($\ref{Bmatrix}$). In \cite{diag_Fred} we show that $\Bmat{a}$ is a lower
 triangular matrix, in which the diagonal terms drop off exponentially,
and 
the terms below the diagonal fall off even faster. This way 
truncating $\Bmat{a}$ is justified, as truncating the
matrix to a finite one introduces only exponentially small errors.

In the local matrix approximation the traces of evolution operator are approximated by
\bea
\left. \tr {\Lop}^n\right|_{\mbox{\tiny saddles}}
=\sum_{p} n_p \sum_{r=1}^{\infty} \delta_{n, n_p r} \tr \Lmat{p}^r
= \sum_{N=0}^{\infty}C_{n,N}\sigma^{2N}
\,,
\label{tracenp}
\eea
where
$\tr \Lmat{p}  = \tr{ \Lmat_{n_p}\Lmat_2\cdots \Lmat_1}$
is the contribution of the $p$ cycle, and the power series in
$\sigma^{2N}$ follows from the expansion ($\ref{BtoL}$) of $\Lmat{a}$
in terms of $\Bmat{a}$. 
The traces of $\Lmat{}^n$ evaluated by ($\ref{BtoL}$) yield a series in
$\sigma^{2N}$,  and
the $\sigma^{2N}$ coefficients $Q_{n,N}$ in the cumulant expansion 
\bea
F=\det(1-z\Lop) = 1-\sum_{n=1}^{\infty}\sum_{N=0}^{\infty}Q_{n,N}z^n\sigma^{2N}
\label{qum}
\eea
are then obtained recursively  from the traces, as in 
($\ref{Fd-cyc-exp-Diag}$):
\bea
Q_{n,N} = \frac{1}{n}\left(C_{n,N}
           - \sum_{k=1}^{n-1}\sum_{l=0}^{N}Q_{k,N-l}C_{n-k,l}
                     \right)
\,.
\label{Q-Sig-expan}
\eea
From the cumulants, by manipulating formal Taylor series, we can calculate 
perturbative corrections to the eigenstates. 

Figure ($\ref{qumfig}$) shows the cumulants obtained in the numerical tests 
 and table (\ref{tablazat}) shows the eigenvalue corrections computed from
 the cumulants.
Both the cumulants and the eigenvalue corrections exhibit a
super-exponential convergence with the truncation cycle length $n$, 
whereas the corrections appear to form a divergent series in the
noise parameter $\sigma$.

\section{Trace formula for noise corrections}
With as many as eight orders of perturbation theory, we
should now turn towards investigating the asymptotic nature
of these perturbative expansions. 
The corrections to the escape rate are calculated via the
cumulant expansion from the traces of the
 evolution operators $\Lop^n$. These traces are responsible for the
asymptotic behavior since they themselves apparently have divergent high order
 noise corrections. Thus in this section
we investigate  the behavior of the late terms in the noise expansion
 series of the traces  Tr$(\Lop^n)$. With the help of an integral
 representation of the operator we shall be able to transform Tr$(\Lop^n)$
 to contour integrals, and then we will evaluate these integrals in
 the saddle point approximation to arrive to a formula 
 analogous to the usual trace formulas arising in quantum chaos.
 
The trace of $\Lop^n$ can be expressed as
\bea
\mbox{Tr}{\cal L}^n=\f{1}{(\sqrt{2\pi}\sigma)^n}\int dx_1dx_2...dx_n
{e}^{-\f{S(\vec x)}{\sigma^2}},
\label{trace}
\eea
where
\bea
S(\vec x)&=&\f{1}{2}\sum_{j=1}^n\left(x_{j+1}-f(x_j)\right)^2, \label{Sdef}\\
\vec x &=& (x_1,\cdots,x_n), \qquad {\rm with}\quad  x_{n+1}\equiv x_1.
\eea

 In order to give a deeper insight on the forthcoming calculations,
we draw a correspondence between our system and a discrete Hamiltonian mechanics,
with the $S$ defined above playing the role of the classical action.
According to ($\ref{Sdef}$),
 the least action principle requires
\bea
x_j-f(x_{j-1})-f'(x_j)(x_{j+1}-f(x_j))=0.
\label{least}
\eea
We define 
\bea
\vec p = (p_1,\cdots,p_n),  \qquad  p_j:=x_j-f(x_{j-1}),
\eea
the quantity corresponding to the momentum in the classical mechanics.
From (\ref{least}) we obtain
\bea
x_{j+1}&=&f(x_j)+p_{j+1}, \\
p_{j+1}&=&\f{p_j}{f'(x_j)},
\eea
which are the equations corresponding to the classical Newtonian equations
 of motion.
The generalized periodic orbits of length $ n$ are those orbits, 
which obey these equations and $x_{n+1}=x_1$,
$p_{n+1}=p_n$. Those generalized periodic orbits which have
non-zero momentum will control the asymptotic behavior
of the corrections to $\mbox{Tr}{\cal L}^n$ 
as we shall demonstrate later.
The original periodic orbits defined by ($\ref{peri1}$),($\ref{peri2}$)
are those with zero momentum. 
 The generalized periodic orbits
with non-zero momentum and the original periodic orbits proliferate
 with growing $n$ as suggested by Fig $\ref{orbits}$.

We introduce an integral representation of the 
noisy kernel, which will be of great use in the later
 calculations:
\bea
\Lop(x,y)&=&\f{1}{\sqrt{2\pi}\sigma}{e}^{-\f{(y-f(x))^2}{2\sigma^2}}=
 \no \\ & &
\f{1}{2\pi}\int dk {e}^{-\f{\sigma^2k^2}{2}+ik(y-f(x))}. \label{theop}
\eea
Using  this new integral representation,
\bea
& &\mbox{Tr}{\cal L}^n=\no \\ 
& &\f{1}{(2\pi)^n}\int dk^ndx^n {e}^{-\f{\sigma^2}{2}
\sum_{j=1}^nk_j^2+i\sum_{j=1}^nk_j(x_{j+1}-f(x_j))},\no \\
\label{ujtrace}
\eea
or equivalently
\bea
& &\mbox{Tr}{\cal L}^n=\no \\ 
& &\f{1}{(2\pi)^n}\int dk^n \int dp^n J_n\left(\vec p\right){e}^{-\f{\sigma^2}{2}
\sum_{j=1}^nk_j^2+i\sum_{j=1}^nk_jp_j}, \label{moment}
\eea
where $J_n(\vec p)$ denotes the Jacobian $D(\vec x)/D(\vec p)$. 
Since
\bea
\f{1}{(2\pi)^n}\int dk^n {e}^{i\sum_{j=1}^n k_jp_j}=\prod_{j=1}^n
\delta(p_j),
\eea
we can reduce ($\ref{moment}$) to 
\bea
\mbox{Tr}{\cal L}^n&=&\int dp^n J_n\left(\vec p\right){e}^{\f{\sigma^2}{2}\Delta_n}
\prod_{j=1}^n\delta(p_j)=\no \\ & &
\left.{e}^{\f{\sigma^2}{2}\Delta_n}J_n\left(\vec p \right)
\right|_{p_j=0}, \label{elegant}
\eea
with $\Delta_n$ denoting the Laplacian
\bea
\Delta_n=\f{\p^2}{\p p_1^2}+\f{\p^2}{\p p_2^2}+...+\f{\p^2}{\p p_n^2}.
\eea
We focus on the Taylor expansion of ($\ref{elegant}$) in the
noise parameter:
\bea
\mbox{Tr}{\cal L}^n&=&\sum_{N=0}^{\infty}\left(\mbox{Tr}{\cal L}^n\right)_N
\sigma^{2N},\\
\left(\mbox{Tr}{\cal L}^n\right)_N&=&
\left.\f{1}{2^N}\f{(\Delta_n)^N}{N!}J_n\left(\vec p\right)\right|_{p_j=0}.
\eea 
The $N$-th power of the Laplacian in the equation above can be written as
\bea
(\Delta_n)^N=
\sum_{j_1,...,j_n=0}^{\infty}\f{N!}{j_1!...j_n!}
\f{\p^{2j_1}}{\p p_1^{2j_1}}...\f{\p^{2j_n}}
{\p p_n^{2j_n}}\delta_{N,\sum_{k=1}^nj_k}, 
\eea
where $\delta_{jl}$ is the Kronecker-delta (making the sum actually finite).
With the help of the multidimensional residue formula from complex calculus
 \cite{comcalc} 
\bea
& &\f{\p^{n_1+...+n_k}f\left(\vec z\right)}{\p z_1^{n_1}...\p z_k^{n_k}}=\no \\ & &
\f{n_1!...n_k!}{(2\pi {i})^k}\oint...\oint\f{
f({\vec \xi})d\xi_1...d\xi_k}{(\xi_1-z_1)^{n_1+1}...(\xi_k-z_k)^{n_k+1}},
\eea
we obtain
\bea
& &\left(\mbox{Tr}{\cal L}^n\right)_N=\f{1}{(2\pi {i})^n2^N}
\sum_{j_1,...,j_n=0}^{\infty}\f{(2j_1)!}{j_1!}...
\f{(2j_{n}))!}{j_{n}!}\no \\ & &\times\delta_{N,\sum_{k=1}^nj_k}
\oint_{c_1}...
\oint_{c_n}\f{J_n\left(\vec p\right)dp_1...dp_n}{p_1^{2j_1+1}...
p_n^{2j_n+1}}.\label{numeri1}
\eea
These contours encircle the $p_j=0$ points positively. 
The integrals can be transformed back to contour integrals in 
the original $x_j$ variables, and the contours will be placed
 around the original periodic orbits of the system defined
by ($\ref{peri1}$--$\ref{peri2}$), since it is these orbits
 which fulfill the $p_j=0$ conditions.
\bea
& &\left(\mbox{Tr}{\cal L}^n\right)_N=\f{1}{(2\pi {i})^n2^N}
\sum_{j_1,...,j_n=0}^{\infty}\f{(2j_1)!}{j_1!}...
\f{(2j_{n}))!}{j_{n}!}\no \\ & &\times\delta_{N,\sum_{k=1}^nj_k}
\oint_{C_1}...
\oint_{C_n}\no \\ & &\f{dx_1...dx_n}{(x_1-f(x_n))^{2j_1+1}...
(x_n-f(x_{n-1}))^{2j_n+1}}.\label{numeri2}
\eea
 From now on we shall restrict
 our calculations to the asymptotic large-$N$ limit. We will replace
 the summations in ($\ref{numeri1}$) by integrals and then use the
 saddle-point method to get a compact formula for 
$(\mbox{Tr}{\cal L}^n)_N$. Then we may consistently
approximate the factorials via the Stirling formula \cite{abramo} as
\bea
\f{(2j_k)!}{j_k!}&\simeq&\f{\left(\f{2j_k}{e}\right)^{2j_k}\sqrt{4\pi j_k}}
{\left(\f{j_k}{e}\right)^{j_k}\sqrt{2\pi j_k}}=2^{2j_k+1/2}j_k^{j_k}
{e}^{-j_k}=\no \\ & &2^{1/2}{e}^{2(\ln 2)j_k+j_k\ln j_k-j_k}. 
\label{StirlingII}
\eea
Using ($\ref{StirlingII}$) and an integral representation of the
 delta function we get
\bea
& &\left(\mbox{Tr}{\cal L}^n\right)_N\simeq\f{2^{\f{n}{2}-N}}{(2\pi {i})^n2\pi}
\sum_{j_1,...,j_n=0}^{\infty}\int dt \oint_{C_1}...\oint_{C_n}dx_1...dx_n
\no \\ & &
\times\exp\left[{i}t(N-\sum_{k=1}^nj_k)+(2\ln 2-1)\sum_{k=1}^nj_k
+\sum_{k=1}^nj_k\ln j_k\right.\no \\& &\left.+\sum_{k=1}^n\ln(x_k-f(x_{k-1}))
(2j_k+1)\right].
\eea
Now we replace $j_k$ with the new variables $y_k=\f{j_k}{N}$ and 
in the asymptotic ($N$ large) limit approximate the summations over $j_k$
 with integrals over $y_k$ as
\bea
& &\left(\mbox{Tr}{\cal L}^n\right)_N\simeq\no \\ & &
\f{2^{\f{n}{2}-N}N^n}{(2\pi {i})^n2\pi}
\int_0^{\infty}dy_1...\int_0^{\infty}dy_n
\int dt \oint_{C_1}...\oint_{C_n}dx_1...dx_n
\no \\ & &
\times\exp\left[{i}t(N-N\sum_{k=1}^ny_k)+N(2\ln 2-1)\sum_{k=1}^ny_k
\right.\no \\ & &\left.+N\sum_{k=1}^ny_k\ln(Ny_k)
+\sum_{k=1}^n\ln(x_k-f(x_{k-1}))(2Ny_k+1)\right].\no \\
\eea
 We evaluate the $y$ integrals with the saddle-point
 method to get 
\bea
\left(\mbox{Tr}{\cal L}^n\right)_N&\simeq&\f{2^{-N+\f{n}{2}}}
{(2\pi)^{\f{n}{2}}{i}^n2\pi}
\int dt \oint_{C_1}...\oint_{C_n}
dx_1...dx_n\no \\ & &
\exp\left[{i}t\left(N+\f{n}{2}\right)-{e}^{it}\f{S}{2}\right].
\label{prefactor}
\eea
Next we implement the saddle-point method to the
integral in $t$ as well, asymptotically resulting in
\bea
\left(\mbox{Tr}{\cal L}^n\right)_N&\simeq&
\f{N^{\f{n-1}{2}}}{2^{2N+\f{1}{2}}(2\pi)^{\f{n+1}{2}}{i}^{n+1}}
\f{(2N)!}{N!}\no \\ & &\times
\int dx^n{e}^{-\left(N+\f{n}{2}\right)\ln(S)}. \label{ucso}
\eea 
The last step is to evaluate the contour integrals in the $x_k$
variables. We deform the contours until the saddle-points are reached
and the contours run along the paths of steepest descent. The
 leading contribution comes from the saddle-points, which fulfill
 the following equation
\bea
\f{1}{S}\left[
x^*_{j}-f(x^*_{j-1})-(x^*_{j+1}-f(x^*_j))f'(x^*_j)\right]=0.
\label{saddle2}
\eea
By comparing ($\ref{saddle2}$) and ($\ref{least}$) one can see that
the saddle-points are all generalized periodic orbits of the system.
 Since the contours ran originally around the 
 orbits with zero momentum, these do not come into account as saddle-points.
The second derivative matrix is
\bea
-\left(N+\f{n}{2}\right)\f{1}{S}D^2 S,
\eea
where $D^2 S$ denotes the second derivative matrix of $S$
\bea
(D^2S)_{ij}=\f{\partial^2S}{\partial x_i\partial x_j}.
\eea
This would be the matrix to deal with if we were taking the 
saddle-point approximation of ($\ref{trace}$) directly.
We reorganize the prefactor in ($\ref{ucso}$) with the use of the 
Stirling formula \cite{abramo}; then the result of the saddle-point integration
comes out as
\bea
& &(\mbox{Tr}{\cal L}^n)_N\simeq\no \\ & &
\sum_{\rm s.p.}\f{N^{\f{n-1}{2}}}{2\pi {i}}
\f{\Gamma(N+\f{1}{2})}{\left(N+\f{n}{2}\right)^{\f{n}{2}}}
\f{S_p^{-N}}{\sqrt{\det D^2 S_p}}. \label{result}
\eea
 For $n=1$ this formula restores the
 result of \cite{asymp} as it should.

Finally we draw attention to the
 close connection between  the generalized periodic orbits of the 
 system and $D^2S$.
The stability matrix of a generalized periodic orbit $p$ is expressed
 as
\bea
M_p&=&M(n)\cdot M(n-1)\cdot M(n-2)\cdots M(1) \\
M(k)&=&\left(\matrix{f'(x_k)-\f{p_k}{(f'(x_k))^2}f''(x_k) & 
\f{1}{f'(x_k)} \cr -\f{p_k}{(f'(x_k))^2}f''(x_k) & \f{1}{f'(x_k)}}
\right)
\eea
The determinant of $D^2S$ can be expressed with the help of
 the stability matrix as
\bea
\det D^2S_p=\det(M_p-1).
\eea
This way we reformulate ($\ref{result}$) as
\bea
(\mbox{Tr}{\cal L}^n)_N\simeq\f{N^{\f{n-1}{2}}}{2\pi}
\f{\Gamma(N+\f{1}{2})}{\left(N+\f{n}{2}\right)^{\f{n}{2}}}
\sum_{\rm g.p.o.}\f{{e}^{-N\ln S_p}}{\sqrt{\det(1-M_p)}}, \label{result2}
\eea
where the summation runs over generalized periodic orbits, with
non-zero momentum. This expression, fully analogous to a trace formula,
 is our main result. 

Finally, we provide a numerical test of our high order estimates.
In \cite{asymp} we developed a contour integral method to calculate
 high order noise corrections to the trace of $\Lop$, and 
obtained a very good agreement between the exact results and a formula
 which is just the approximation of ($\ref{result}$) in the $n=1$ case.
Now we can similarly verify the high order noise
corrections to the trace of $\Lop^2$ .
 Fig. $\ref{abra}$ plots, as a function of $N$, 
the ratios of the approximations for
$(\mbox{Tr}\Lop)_N$ and $(\mbox{Tr}\Lop^2)_N$ obtained from
 ($\ref{result}$) to the
exact results ($\ref{numeri2}$) computed using direct numerical evaluations of
 the contour integrals therein. 
(For higher powers of $\Lop$,
 the proliferation of the orbits forces the contours to run 
close to singular points, making efficiently precise numerical integrations
extremely time-consuming).

In summary, we have  studied  the evolution operator for a discrete 
Langevin equation with a strongly hyperbolic 
classical dynamics and a Gaussian noise distribution. Motivated by
 the divergent growth of the high order eigenvalue corrections
 of the evolution operator, we investigated the behavior of the
 high order noise corrections to the traces of $\Lop^n$.
 Using an 
integral representation of the evolution operator $\Lop$, we found that
 the asymptotic behavior of the  corrections to 
the trace of $\Lop^n$ is governed by sub-dominant
quantities 
previously neglected in the perturbative expansion, and
 a full-fledged trace formula can be derived for the late
 terms in the noise expansion series of the trace of $\Lop^n$.

G.V. and G.P. gratefully acknowledge the financial support of
the Hungarian Ministry of Education, EC Human Potential Programme
, OTKA T25866.
G.P., G.V. and A.V. were also partially supported by the French Minist\`ere des
 Affaires {\'E}trang\`eres.

\begin{figure}[hbt]
%\centerline{\strut\psfig{figure=tmap.eps,height=6cm}}
\caption{ The map ($7$)
 on the [0,1] interval.
\label{figmap}}
\end{figure}

\begin{figure}[hbt]
%\centerline{\psfig{figure=qum1.eps,height=8cm}}
\caption{The perturbative corrections 
 to the cumulants $Q_{n,N}$ plotted as a function
of cycle length $n$ (for perturbation orders $N=0,2,4,6,8$); all exhibit
super-exponential convergence in $n$.}
\label{qumfig}
\end{figure}

%\begin{figure}[hbt]
%\centerline{\strut\psfig{figure=fratio0.ps,height=8cm}}
%\caption{Ratio of noise corrections $\f{I_{2 n + 2}}{I_{2 n}}$ for orbit 0. }
%\label{ratioZero}
%\end{figure}

%\begin{figure}[hbt]
%\centerline{\strut\psfig{figure=fratio1.ps,height=8cm}
%\caption{Ratio of noise corrections $\f{I_{2 n + 2}}{I_{2 n}}$ for orbit 1 }
%\label{ratioOne}
%\end{figure}

\begin{figure}[hbt]
%\centerline{\strut\psfig{figure=neworbits2.eps,height=8cm}}
\caption{ The shortest original and generalized periodic orbits of the map
 ($7$). Squares
 mark original periodic points, dots mark generalized periodic points.
 Large symbols indicate orbits of length one, small symbols indicate orbits
 of length two.}
\label{orbits}
\end{figure}

\begin{figure}[hbt]
%\centerline{\strut\psfig{figure=trace.eps,height=8cm}}
\caption{ The ratios of $(\mbox{Tr}{\cal L})_N$ and $(\mbox{Tr}{\cal L}^2)_N$
 as calculated via
the asymptotic formula (55) to their values computed by 
numerical integration of formula (46).}
\label{abra}
\end{figure}
\newpage
\centerline{\strut\psfig{figure=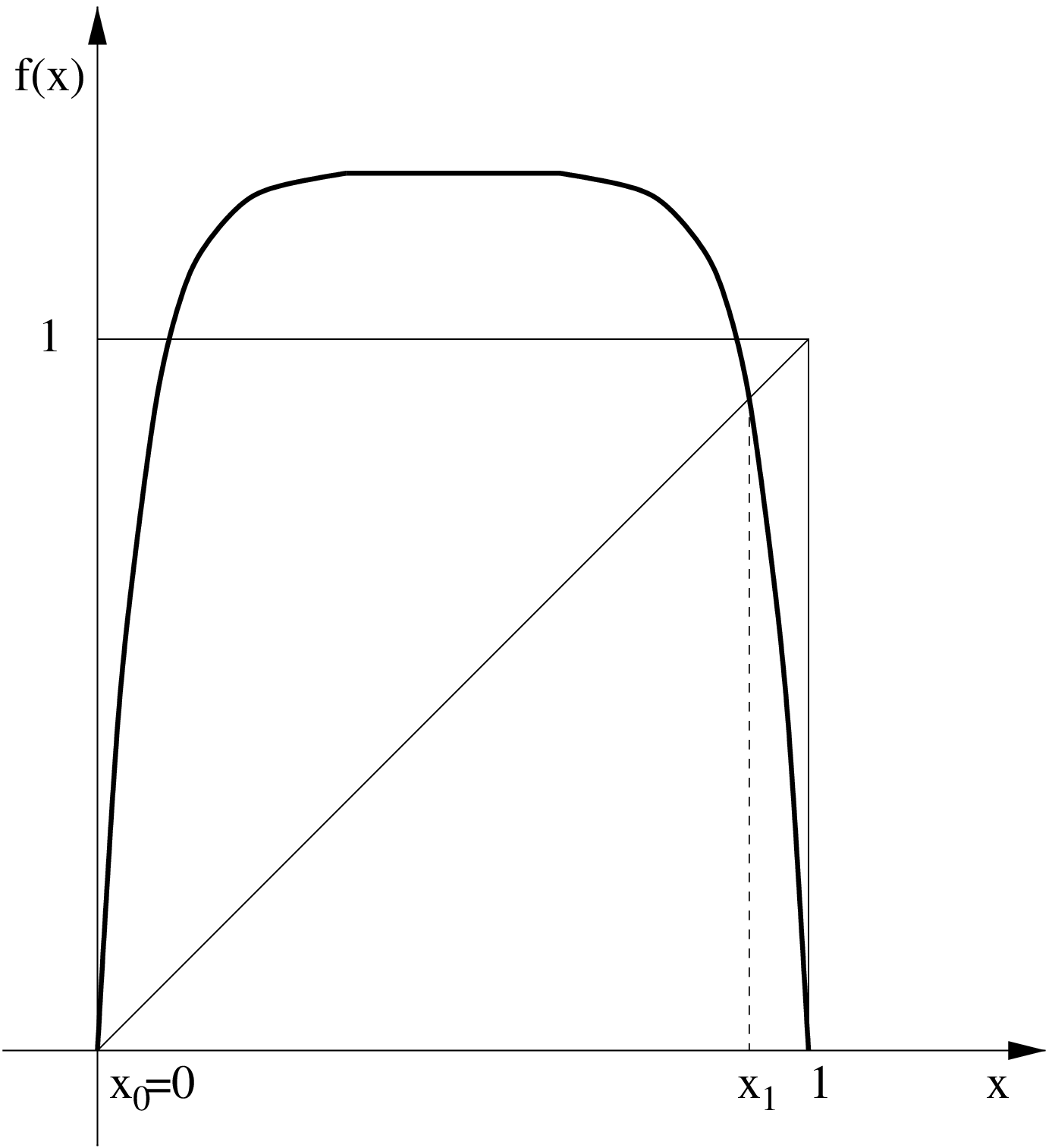,height=10cm}}
\center{FIG.1.}
\newpage
\centerline{\psfig{figure=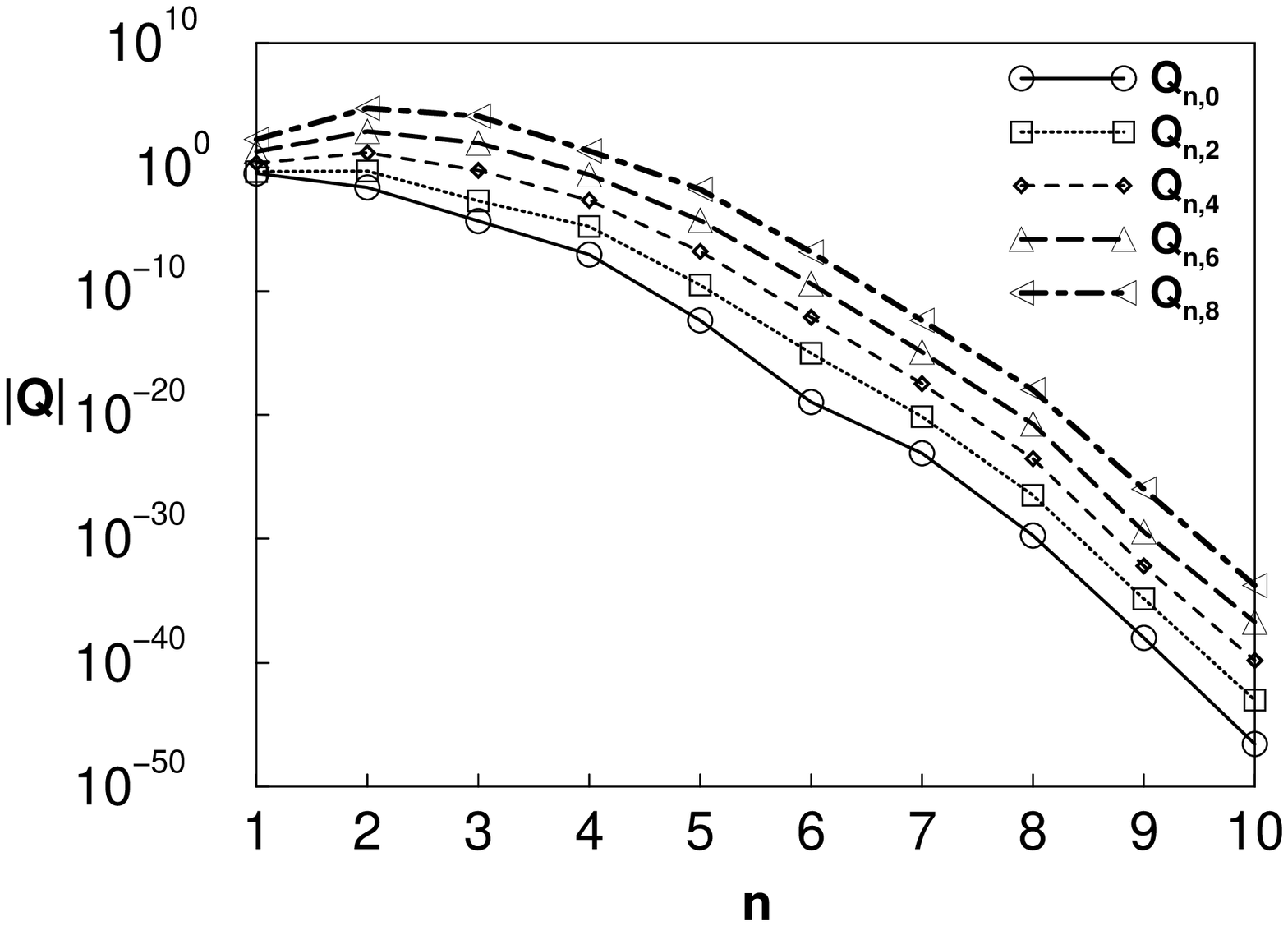,height=10cm}}
\center{FIG.2.}
\newpage
\centerline{\strut\psfig{figure=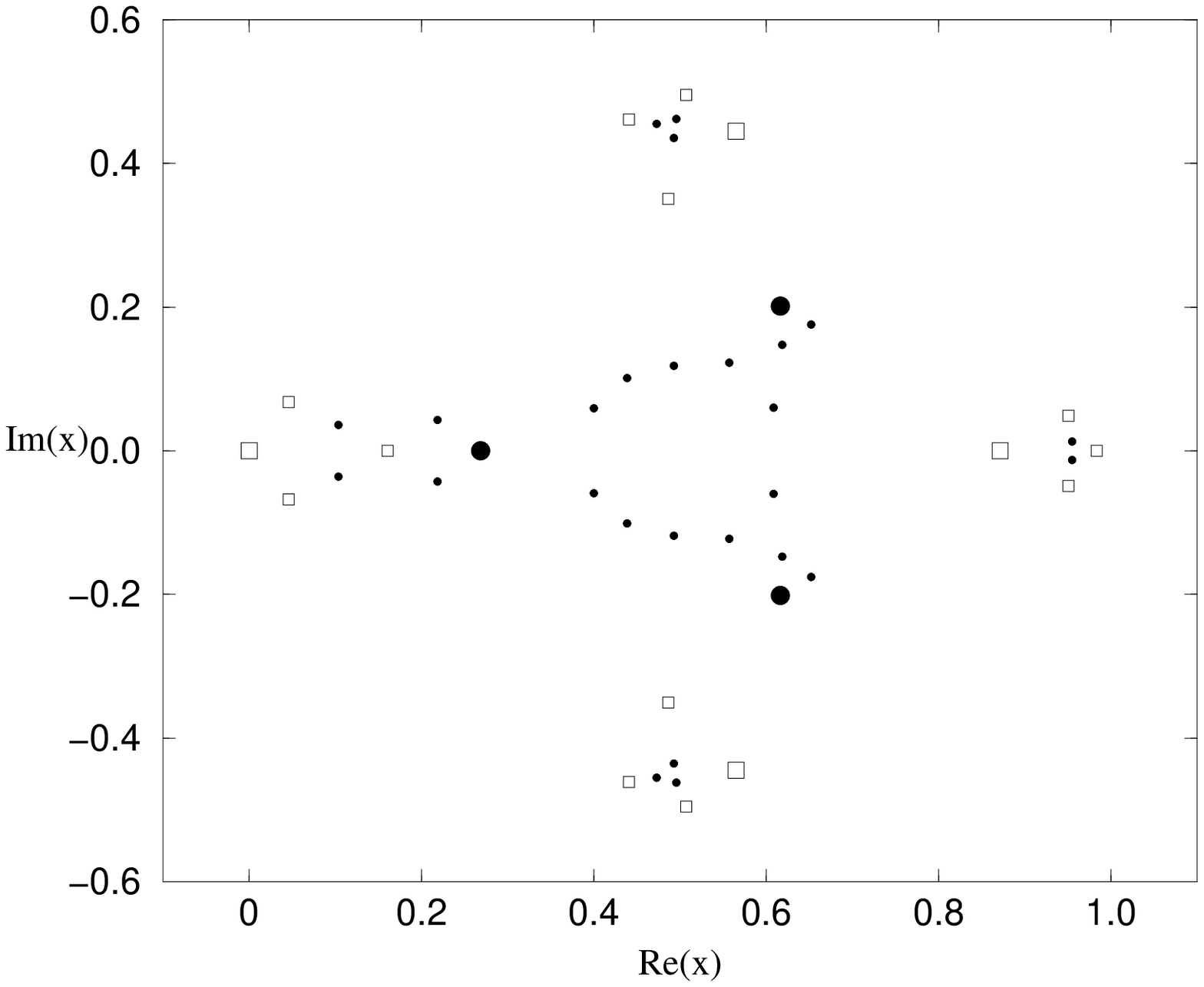,height=10cm}}
\center{FIG.3.}
\newpage
\centerline{\strut\psfig{figure=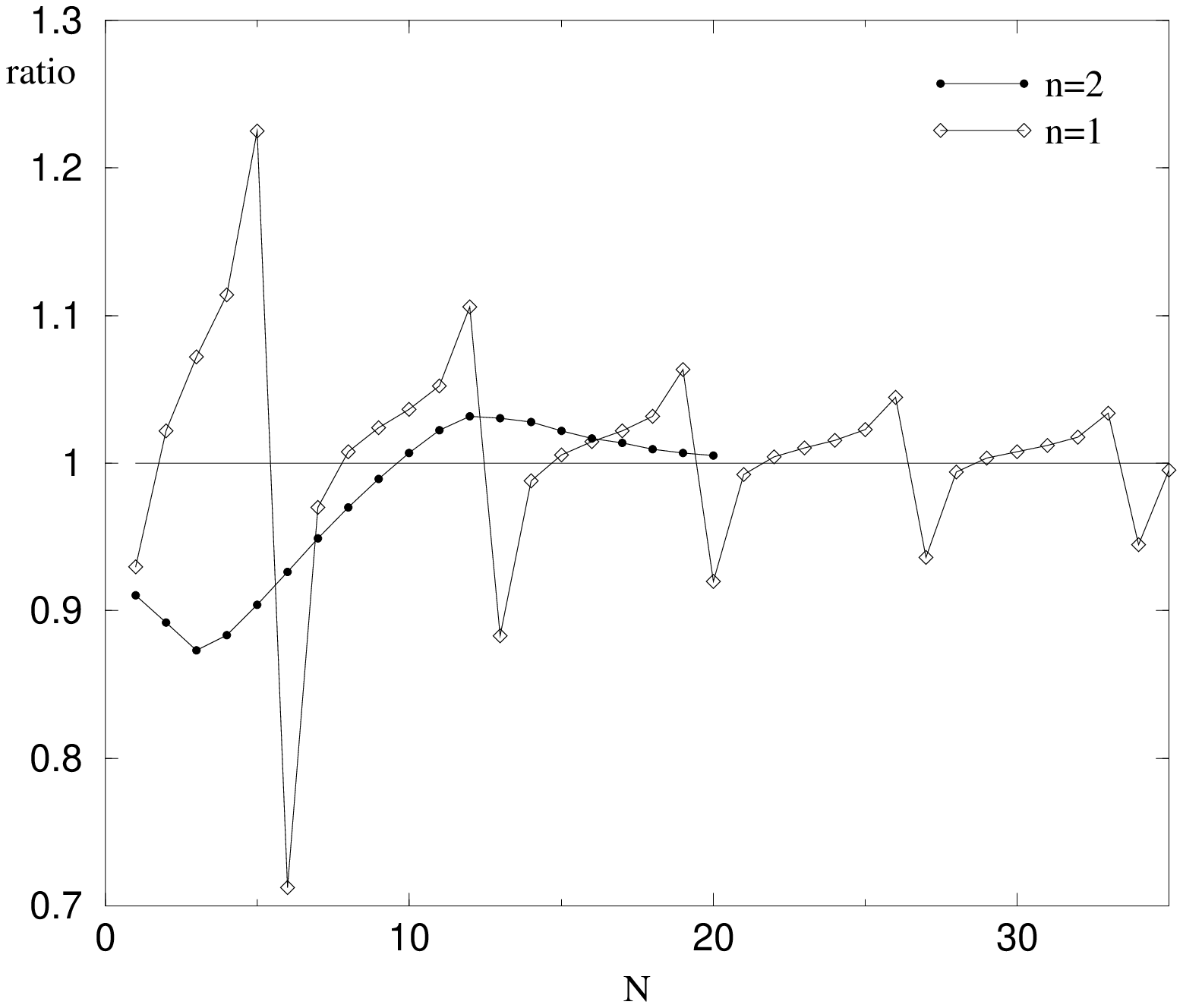,height=10cm}}
\center{FIG.4.}
\newpage

\begin{table*}
\begin{tabular}{clllll}
%\hline
$n$&$\nu_0$&$\nu_2$&$\nu_4$&$\nu_6$&$\nu_8$\\\hline
1&0.308&0.42&~2.2                                    &17.4             &168.0    \\
2&0.37140&1.422&32.97                                &1573.3           &112699.9 \\
3&0.3711096&1.43555&36.326                           &2072.9           &189029.0 \\
4&0.371110995255&1.435811262&36.3583777              &2076.479         &189298.8 \\
5&0.371110995234863&1.43581124819737&36.35837123374  &2076.4770492     &189298.12802\\
6&0.371110995234863&1.43581124819749&36.358371233836 &2076.47704933320 &189298.128042526\\
\hline
%6&0.371110995234863& 0.7179 & 1.5149 &2.88399 &4.69489404867376\\
%\hline
\end{tabular}
\caption{Significant digits of the leading deterministic eigenvalue
$\nu_0$, and of its
$\sigma^2, \cdots,  \sigma^{8}$ perturbative coefficients $\nu_2,\cdots,\nu_8$,
 calculated from the cumulant expansion of the spectral determinant
as a function of the cycle truncation length $n$.
Note the super-exponential convergence of all coefficients. 
\label{tablazat}
}
\end{table*}
%REFERENCES -------------------------------------------------------

\end{document}